\documentstyle[preprint,revtex]{aps}
\begin{document}
\draft

\preprint{WUGRAV 92-14}
\begin{title}
Spin effects in the inspiral of coalescing compact binaries
\end{title}

\author{Lawrence E. Kidder and Clifford M. Will}

\begin{instit}
McDonnell Center for the Space Sciences, Department of
Physics\\
Washington University, St. Louis, Missouri 63130
\end{instit}

\author{Alan G. Wiseman}

\begin{instit}
Department of Physics and Astronomy\\
Northwestern University, Evanston, Illinois 60208
\end{instit}

\begin{abstract}

We derive the contributions of spin-orbit and spin-spin coupling to
the gravitational radiation from coalescing binary systems of spinning
compact objects.  We calculate spin effects in the symmetric, trace-free
radiative multipoles that determine the gravitational waveform,
and the rate of energy loss.  Assuming a balance between energy
radiated and orbital energy lost, we determine the spin effects
in the evolution of the
orbital frequency and orbital radius.  Assuming that a laser interferometric
gravitational observatory can track the gravitational-wave frequency (twice the
orbital frequency) as it sweeps through its
sensitive bandwidth between about 10 Hz and one kHz,
we estimate the accuracy with which the spins of
the component bodies can be determined from the gravitational-wave signal.
\end{abstract}


\narrowtext

\section{Introduction}

The ability of laser interferometric gravitational-wave (GW) detectors such
as the US LIGO to extract useful astrophysical information about
coalescing binary systems of compact objects from the observed
gravitational-wave signals depends upon highly accurate theoretical
models for the decaying orbital evolution.  Detection
and study of the characteristic ``chirp'' waveform emitted by such
systems involves a matched filtering technique using a theoretical
template that is a function of parameters of the source
\cite{filters}.
As has been pointed out by Cutler {\it et al.} \cite{jugger}, matching
the theoretical variation of the GW {\it frequency} with the data
as the frequency sweeps through
the detector bandwidth between 10 and 1000 Hz
may be the most
promising way to obtain estimates of
the masses and spins of the bodies.  To this end, one
needs a theoretical formula for the evolution of the GW frequency that
includes all dependence on the two masses and spins,
and that is sufficiently accurate that errors in the model
lead to cumulative phase errors no larger than $2\pi$ radians
over the potentially thousands of cycles observed in the frequency
bandwidth of interest (larger phase errors substantially reduce the
signal-to-noise of the matched filter).

Cutler {\it et al.} \cite{jugger} have suggested that a model for
the frequency evolution that is based on a post-Newtonian
approximation may be inadequate unless the approximation is carried to
extremely high orders beyond the lowest-order Newtonian
and post-Newtonian corrections that
have been derived to date \cite{wagwill,linc}.  In addition, they
point out that the effects of spin-orbit and spin-spin coupling of the
bodies have not been derived, even in the first post-Newtonian
approximation.  It is the purpose of this paper to remedy the latter
situation.

We derive the spin-orbit and spin-spin contributions to
the symmetric trace-free (STF) multipoles that enter the gravitational
waveform for two-body systems,
using a multipole formalism developed by Blanchet, Damour
and Iyer (BDI) \cite{bd,di}.  In addition to explicit spin terms,
the multipole formalism involves several
time-derivatives of multipole moments that lead to expressions involving the
two-body acceleration; to evaluate these moments consistently, we
use post-Newtonian equations of motion that
also include spin terms.
Using these multipoles, we then evaluate the rate of energy
loss from the binary system.  Specializing to a circular orbit, we
calculate the rate of decrease of orbital radius and the rate of
change of orbital frequency, assuming energy balance
and using an expression for energy that
includes spin terms.
Finally, we estimate the size of spin-orbit
and spin-spin
effects for coalescing binary systems of neutron stars and/or black
holes on the accumulation of GW phase in a
LIGO-type detector.

\section{Equations of motion}

Equations of motion for bodies of arbitrary mass and spin
have been developed by numerous authors (for reviews
and references see \cite{barker,damour300,brumberg}).
For our purposes, only the Newtonian, spin-orbit (SO) and spin-spin
(SS) terms
will be needed; post- and post-post-Newtonian non-spin terms have been
considered elsewhere \cite{linc}.
Although the inclusion of spin implies that we are treating extended
bodies, we ignore tidal and quadrupole-coupling effects (even though
rotationally induced quadrupole effects are proportional to (spin)$^2$);
for binary
systems containing neutron stars or black holes, these are expected to be small
until the very latest stage of inspiral and coalescence, except
possibly for rapidly rotating Kerr black holes \cite{bildsten}.

By eliminating the center of mass of the system,
we convert the two-body equations of motion to an
effective one-body equation of motion given by
($G=c=1$)

\begin{equation}
{\bf a} \equiv {\bf a_1} - {\bf a_2} = - {m \over r^2} {\bf \hat n} +
{\bf a_{SO}} + {\bf a_{SS}},\label{eom}
\end{equation}
where
\begin{mathletters}
\begin{equation}
{\bf a_{SO}} = {1 \over r^3} \left\{ 6 {\bf \hat
  n} \left[ \left( {\bf \hat n \times v} \right) {\bf \cdot} \left( {\bf \zeta}
 + {\bf \xi} \right) \right] - \left[ {\bf v \times} \left( 4{\bf \zeta} + 3
  {\bf \xi} \right) \right] + 3 \dot r \left[ {\bf \hat n \times}
  \left( 2{\bf \zeta} + {\bf \xi} \right) \right] \right\},\label{eomso}
\end{equation}
\begin{equation}
{\bf a_{SS}} = -{3 \over \mu r^4} \left\{ {\bf \hat n} \left( {\bf S_1 \cdot
  S_2} \right) + {\bf S_1} \left( {\bf \hat n \cdot S_2} \right) + {\bf S_2}
  \left( {\bf \hat n \cdot S_1} \right) - 5 {\bf \hat n} \left( {\bf \hat n
  \cdot S_1} \right) \left( {\bf \hat n \cdot S_2} \right) \right\},
\label{eomss}
\end{equation}
\end{mathletters}
where
$m=m_1+m_2$,
$\mu = m_1m_2/m$,
$r = |{\bf x}|$,
${\bf x} = {\bf x_1} - {\bf x_2}$,
${\bf \hat n} = {\bf x} / r$,
${\bf v} = {\bf v_1} - {\bf v_2}$,
${\bf \zeta} = {\bf S_1} + {\bf S_2}$,
${\bf \xi} = (m_2/m_1){\bf S_1} + (m_1/m_2){\bf S_2} $,
and $\dot r = ({\bf \hat n \cdot v})$.

It is useful to note that
spin-orbit and spin-spin effects are of order $(R/r)v\bar v$ and
$(R/r)^2 {\bar v}^2$, respectively, compared to the Newtonian
acceleration, where $R$ and $\bar v$ denote the characteristic radius
and rotation velocity of each body; these terms thus are formally
of first-post-Newtonian order.  However, for compact bodies, $R$
is of order of $m$ or a few times $m$, and $\bar v$ could be of order
unity, so that in practice
these terms can be considered to be of post$^{3/2}$-Newtonian and
post-post-Newtonian order, respectively (indeed they are so denoted in
\cite{jugger}).

Equations for the precession of the spins caused by spin-orbit
(geodetic precession) and spin-spin (Lense-Thirring) couplings
\cite{barker,damour300,brumberg} can also
be written down in terms of relative coordinates.  The
relevant equation for our purposes is
\begin{equation}
{\bf \dot \zeta} = {1 \over r^3} {\bf L_N \times} \left( 2 {\bf \zeta} +
  {3 \over 2} {\bf \xi} \right) + {3 \over r^3} \left[ \left( {\bf \hat n
  \cdot S_2} \right) \left( {\bf \hat n \times S_1} \right) + \left( {\bf
  \hat n \cdot S_1} \right) \left( {\bf \hat n \times S_2} \right) \right].
\label{precess}
\end{equation}
where ${\bf L}_N \equiv \mu {\bf x} \times {\bf v}$
denotes the Newtonian orbital angular momentum.
Note that the precession of the spins is one order higher than the
spins themselves.  Therefore we may neglect this precession
when evolving the equations to lowest order; Eq. (\ref{precess}) is only needed
to
verify that total angular momentum is conserved for the system, apart
from that radiated away \cite{sussman}.

These equations of motion can be derived from
a generalized Lagrangian which is a function of the relative position,
velocity, and acceleration,
given by
\begin{equation}
{\cal L} = {1 \over 2} \mu v^2 + \mu{m \over r} +
{\cal L}_{SO} + {\cal L}_{SS},\label{lagrange}
\end{equation}
where
\begin{mathletters}
\begin{equation}
{\cal L}_{SO} = {1 \over 2} {\mu \over
  m} \left[ {\bf v \cdot} \left( {\bf a \times \xi} \right) \right] + 2
  {\mu \over r^3} \left\{ {\bf v \cdot} \left[ {\bf x \times} \left( {\bf
  \zeta} + {\bf \xi} \right) \right] \right\},
\label{lagrangeso}
\end{equation}
\begin{equation}
{\cal L}_{SS} = {1 \over r^3} \left\{ \left( {\bf S_1 \cdot S_2}
  \right) - 3 \left( {\bf
  \hat n \cdot S_1} \right) \left( {\bf \hat n \cdot S_2} \right) \right\}.
\label{lagrangess}
\end{equation}
\end{mathletters}
Here the Euler-Lagrange equations are
${\partial {\cal L} / \partial x^i} - dp_i /dt = 0$,
where
$p^i = {\partial {\cal L} / \partial v^i}  - \dot s^i$, and
$s^i = {\partial {\cal L} / \partial a^i}$.
It is understood that wherever the acceleration appears in
higher-order terms in the Euler-Lagrange equation,
one substitutes the lower-order equation of motion (see
\cite{lagrange} for discussion of acceleration-dependent Lagrangians).

The relative Lagrangian is invariant with respect to time translations
so that there exists a conserved
energy, given by $E = p^iv^i + s^ia^i - {\cal L}$.
Evaluating this expression we obtain
\begin{equation}
E = {1 \over 2} \mu v^2 - \mu{m \over r} +
E_{SO} + E_{SS},\label{energy}
\end{equation}
where
\begin{mathletters}
\begin{equation}
E_{SO} = {\mu \over r^2} \left\{ \left(
   {\bf \hat n \times v} \right) {\bf \cdot \xi} \right\},
\label{energyso}
\end{equation}
\begin{equation}
E_{SS} = {1 \over r^3} \left\{ 3 \left( {\bf \hat n \cdot S_1} \right)
  \left( {\bf \hat n \cdot S_2} \right) - \left( {\bf S_1 \cdot S_2}
  \right) \right\}. \label{energyss}
\end{equation}
\end{mathletters}
We can define the total angular momentum as
${\bf L} = {\bf \zeta} + \left( {\bf x \times p} \right)
+ \left( {\bf v \times s} \right)$, with the result
\begin{equation}
{\bf L} = {\bf \zeta} + {\bf L}_N +
{\bf L}_{SO} ,\label{totalL}
\end{equation}
where
\begin{equation}
{\bf L}_{SO} = {\mu \over m} \left\{
  {m \over r} \left[ {\bf \hat n \times} \left( {\bf \hat n \times} \left[
  2 {\bf \zeta} + {\bf \xi} \right] \right) \right] - {1 \over 2} \left[
  {\bf v \times} \left( {\bf v \times \xi} \right) \right] \right\},
\end{equation}
where there is no spin-spin contribution to $\bf L$.
Using the equations of motion and spin precession, Eqs. (\ref{eom}) and
(\ref{precess}),
it is straightforward to show explicitly that, to post-Newtonian
order, $\dot E = \dot {\bf L} = 0$.

\section{Evaluation of BDI multipoles}

The radiative energy loss of the system can be expressed in terms of
symmetric and trace-free (STF) radiative multipole
moments (see \cite{rmp} for a review).
For the accuracy we require the energy loss rate is given by
\begin{equation}
{dE \over dt} = - {1 \over 5} \left\{ \mathop{I_{ij}}^{(3)}
 \mathop{I_{ij}}^{(3)} + {5 \over 189} \mathop{I_{ijk}}^{(4)}
 \mathop{I_{ijk}}^{(4)} + {16 \over 9} \mathop{J_{ij}}^{(3)}
\mathop{J_{ij}}^{(3)} \right\}. \label{edot}
\end{equation}
where $I_{ij}$ and $I_{ijk}$ are the STF ``mass'' quadrupole and octopole
moments, and $J_{ij}$ is the STF ``current'' quadrupole moment,
and $(n)$ over each moment denotes
time derivatives.

We evaluate the radiative multipoles using the formalism developed by
Blanchet, Damour and Iyer \cite{bd,di} (BDI).
We restrict ourselves to the case of two well-separated, approximately
spherically symmetric, rotating compact objects
whose structure is given by that of a perfect fluid.  Following our
previous post-Newtonian approach \cite{wagwill,agw92}, we choose the
following definition for the center of mass of each body
\begin{mathletters}
\begin{equation}
 x_A^i = {1 \over m_A} \int\limits_A x^i \rho^* \left( {\bf x} \right)
  \left[ 1 + {1 \over 2} \bar v_A^2 + \Pi - {1 \over 2} \bar U_A \right]
  d^3x  ,
\label{cm}
\end{equation}
\begin{equation}
m_A = \int\limits_A \rho^* \left( {\bf x} \right) \left[ 1 + {1 \over 2}
\bar v_A^2 + \Pi - {1 \over 2} \bar U_A \right] d^3x ,
\end{equation}
\end{mathletters}
where $\rho^*=\rho(1+{1 \over 2}v^2 + 3U)$ is the so-called
``conserved density'' \cite{tegp}, with $\rho$ the local mass
density, $v$ the velocity, and $U$ the Newtonian gravitational potential;
$\Pi$ is the specific internal energy density,
$ \bar v_A^i = v^i - v_A^i$, $v_A^i = dx_A^i /dt$,
and $ \bar U_A $ is the Newtonian potential produced by the $A$-th body
itself.

However, when spin effects are to be included, there is a subtle
difference between the center of mass defined above, and the center of
mass used in the equations of moton (\ref{eom}).  The latter defines the
center of mass world line ${x_A}^{\mu}$
of each body using a so-called ``spin supplementary
condition'' (SSC), given by ${S_A}^{\mu\nu} {u_A}_{\nu} = 0$, where
$u_A^\mu$ is the four-velocity of the center-of-mass world line, and
\begin{equation}
{S_A}^{\mu\nu} \equiv 2 \int (x^{[\mu} - {x_A}^{[\mu}) \tau^{\nu ] 0}
 d^3x ,
\end{equation}
where $\tau^{\mu\nu}$ denotes the stress-energy tensor of matter plus
gravitational fields, satisfying ${\tau^{\mu\nu}}_{,\nu}=0$, and
square brackets around indices denote antisymmetrization.
Note that the spin $\bf S$
of each body is defined by
$S_A^i = {1 \over 2} \epsilon_{ijk}S_A^{jk}$.

It is then simple to show that, evaluating the BDI multipoles using the center
of mass definition Eq. (\ref{cm}), and then making the transformation
\begin{equation}
x_A^i \longrightarrow x_A^i + {1 \over 2m_A} \left( {\bf v_A \times S_A}
  \right)^i , \label{cmshift}
\end{equation}
we can convert all expressions to the center of mass defined by the
SSC of the equations of motion.  Because the correction is of
post-Newtonian order, it needs to be made only in the Newtonian-order
multipoles.

A useful check of the consistency of our approach comes from
evaluating the mass dipole moment of the system, given by Eq. (A16b)
of Ref. \cite{di}, and making
the transformation (\ref{cmshift}).  The result is
$I^i = \sum\limits_A \left\{ m_A x_A^i + \left( {\bf v_A \times
S_A} \right)^i \right\}$.
The original two-body equations of motion from which Eq. (\ref{eom})
is derived \cite{damour300}
then imply that $\ddot I^i = 0$, as expected
(uniform motion of the system's center of mass).

Evaluating the other multipoles, transforming them to the SSC of our
equations of motion using Eq. (\ref{cmshift}),
and then transforming them into relative
coordinates, we obtain
\begin{mathletters}
\begin{equation}
I^{ij} = \mu \left( x^i x^j \right)^{STF} + {8 \over 3} \eta \left[ x^i
  \left( {\bf v \times \xi} \right)^j \right]^{STF} - {4 \over 3} \eta
  \left[ v^i \left( {\bf x \times \xi} \right)^j \right]^{STF} ,
\end{equation}
\begin{equation}
I^{ijk} = \mu \left( x^i x^j x^k \right)^{STF},
\end{equation}
\begin{equation}
J^{ij} = - \mu \left( {\delta m \over m} \right) \left[ x^i \left( {\bf x
  \times v} \right) ^j \right]^{STF} + {3 \over 2} \left( x^i \sigma^j
  \right)^{STF},
\end{equation}
\end{mathletters}
where $\delta m = m_1 - m_2$, $\eta=\mu/m$, and
${\bf \sigma} = (m_2/m){\bf S_1} - (m_1/m){\bf S_2}$.  Note that, to
post-Newtonian order, there are no explicit spin-spin contributions to the
multipoles.

\section{Energy loss and inspiral of circular orbits}

Taking time derivatives of Eqs. (14), substituting the equations
of motion where appropriate, and substituting the results into
Eq. (\ref{edot}), we obtain the energy loss rate
\begin{eqnarray}
{dE \over dt} = && - {8 \over 15} {m^2 \mu^2 \over r^4} \left\{ 12v^2 - 11
\dot r ^2 \right.  \nonumber\\
&& \left. +{ {\bf \hat n \times v}  \over mr} {\bf \cdot} \left[
{\bf \zeta} \left( 27\dot r ^2 -37v^2 - 12 {m \over r}
\right)
+ {\bf \xi} \left( 51\dot r ^2 -43v^2 + 4{m \over r} \right) \right]
\right.
\nonumber\\ && \left. + {1 \over 2m \mu r^2} \left[
3  ({\bf S_1 \cdot S_2})  \left( 47 v^2- 55 \dot r ^2
\right)
-  3 (
{\bf \hat n \cdot S_1} {\bf \hat n \cdot S_2} )
\left( 168 v^2 - 269 \dot r ^2 \right) \right. \right.  \nonumber\\
&& \left. \left.
+ 71 ( {\bf v \cdot S_1} {\bf v \cdot S_2} )
- 171 \dot r \left( {\bf v \cdot S_1} {\bf \hat n \cdot
S_2} + {\bf \hat n \cdot S_1}
  {\bf v \cdot S_2} \right) \right] \right\} . \label{eloss}
\end{eqnarray}

We now restrict ourselves to the case of nearly circular orbits, by which
we mean orbits whose inspiral timescale due to gravitational
radiation energy loss is long compared to an orbital timescale, and in
which $r = {\rm constant} +$ small, periodic perturbations due to spin
couplings (when the spins are orthogonal to the orbital plane, the
spin perturbations are constant).  To simplify the
discussion, we then take an angular average of all quantities over an
orbit.
We then have $\ddot r = \dot r = {\bf n} \cdot {\bf v} = 0$,
$v^2 = r^2\Omega^2$, where $\Omega$ is the angular frequency,
${\bf L_N} = \mu r^2\Omega {\bf \hat L}$, where ${\bf \hat L}$ is a
unit vector orthogonal to the orbital plane.  The equation
$ {\bf \hat n \cdot a} = \ddot r - r \Omega ^2$
yields the following orbit-averaged relationship for a circular orbit
\begin{equation}
r = \Omega^{-2/3} m^{1/3} \left\{ 1 - {1 \over 3}
{\Omega \over m} {\bf \hat L \cdot}
\left( 2 {\bf \zeta} + 3 {\bf \xi} \right) - {1 \over 2} {\Omega^{4/3} \over
\mu
  m^{5/3}} \left( {\bf S_1 \cdot S_2} - 3  {\bf \hat
  L \cdot S_1} {\bf \hat L \cdot S_2} \right)
  \right\} . \label{circradius}
\end{equation}

These conditions lead to the following results for the energy and
energy loss rate for a circular orbit, averaged over an orbit,
expressed in terms of
$\Omega$:
\begin{mathletters}
\begin{equation}
E\left(\Omega\right) = - {1 \over 2} \mu \left( m\Omega \right)^{2/3}
 \left\{ 1 + {\Omega \over m} {\bf \hat L \cdot} \left( {8 \over 3} {\bf \zeta}
  + 2 {\bf \xi} \right)  + {\Omega^{4/3} \over \mu
  m^{5/3}} \left( {\bf S_1 \cdot S_2}  - 3 {\bf \hat
  L \cdot S_1} {\bf \hat L \cdot S_2} \right) \right\},\label{Ecirc}
\end{equation}
\begin{eqnarray}
{dE \over dt} \left(\Omega\right) = &&- {32 \over 5} \eta^2 \left(
 m\Omega \right)^{10/3}
 \left\{ 1 - {1 \over 4} {\Omega \over m} {\bf \hat L \cdot} \left(11{\bf
\zeta}
  + 5 {\bf \xi} \right) \right. \nonumber\\
&& \left. - {1 \over 48} { \Omega ^ {4/3} \over \mu m^{5/3}}
 \left( 103 \, {\bf S_1 \cdot S_2}
- 289 \, {\bf \hat L \cdot S_1} {\bf \hat
  L \cdot S_2} \right) \right\} .\label{edotcirc}
\end{eqnarray}
\end{mathletters}
The result for the evolution of orbital frequency is
\begin{eqnarray}
{\Omega^2 \over \dot \Omega} = && {\Omega^2 {dE / d\Omega} \over {dE /
dt}} = {5 \over 96} {m \over \mu} \left( m \Omega \right)^{-5/3} \left\{ 1 +
{\Omega \over m} {\bf \hat L \cdot} \left( {113 \over 12} {\bf \zeta} + {25
\over 4} {\bf \xi} \right)  \right. \nonumber\\
&& \left. + {1 \over 48} {\Omega ^ {4/3} \over \mu m^{5/3}}
\left( 247\, {\bf S_1 \cdot S_2}  - 721 \,
{\bf \hat L \cdot S_1} {\bf \hat L \cdot S_2}
\right) \right\}. \label{omegadot}
\end{eqnarray}

The energy and loss rate can also be expressed exclusively in terms of
$r$ using Eq. (\ref{circradius}),
and an equation for the rate of inspiral obtained.  The result is
\begin{eqnarray}
\dot r = && - {64 \over 5} \eta \left({m \over r}\right)^3 \left\{ 1 -
{7 \over 12} {(m/r)^{3/2} \over m^2} {\bf \hat L \cdot} \left( 19 {\bf \zeta} +
15 {\bf \xi} \right)  \right. \nonumber\\
&& \left. - {5 \over 48} {(m/r)^2 \over \mu m^3}
\left( 59\, {\bf S_1 \cdot S_2}  - 173 \,
{\bf \hat L \cdot S_1} {\bf \hat L \cdot S_2}
\right) \right\}.\label{rdot}
\end{eqnarray}

\section{Discussion of Results}

A signal template whose frequency evolution is given by Eq. (\ref{omegadot})
can be said to match the signal if the accumulated phase in GW from
the time the signal enters the detectors' sensitive bandwidth to the
time it leaves it differs from that of the template by less than
$2\pi$ radians.  Thus one can obtain a crude estimate of the accuracy
with which a parameter characterizing the template can be determined
by finding that change in the parameter that leads to a change of
$2\pi$ in the accumulated phase \cite{jugger}.
Such estimates are only crude and
probably optimistic, because they do not take into account
signal-to-noise issues or correlations among multiple parameters.

The accumulated phase in gravitational waves is given by
\begin{equation}
\Phi_{GW} \equiv 2\pi {\int_{t_i}}^{t_f} f dt = 2 {\int_{\Omega_i}}^{\Omega_f}
 (\Omega^2/\dot\Omega)d\Omega/\Omega  , \label{phase}
\end{equation}
where $f=\Omega /\pi$ is the GW signal frequency.  Integrating the SO
and SS terms in Eq. (\ref{omegadot}), and ignoring precession of the spins or
of the orbital angular momentum, we can estimate the accuracy of
determination of the spin parameters.  For example, for two equal-mass
neutron stars, we find, for the spin-orbit terms,
\begin{equation}
{\Delta\zeta_z \over \zeta_z} \simeq 0.38 \, \left({m \over 2.8
M_\odot}\right)^{5/3}
\left({f_{\rm in} \over 30 {\rm Hz}}\right)^{2/3}
\left({10 {\rm km} \over r_{NS}}\right)^2
\left({P_{NS} \over 10 {\rm ms}}\right) ,\label{estimate}
\end{equation}
where $r_{NS}$ and $P_{NS}$ are the neutron-star radius and rotation
period (suitably averaged over the two stars), and $f_{\rm in}$
is the GW frequency entering the detector bandwidth.  For a 10
$M_\odot$ black hole with spin $S_1 \equiv m_1a_1$, where $a_1<m_1$ is the
Kerr parameter, and a 1.4 $M_\odot$ neutron star, the quantity $a_1/m_1$
be determined to an error
$\Delta (a_1/m_1) \simeq 0.016$ for $a_1/m_1 > 0.03$, while for $a_1/m_1 \ll
0.03$, the spin of the companion neutron star cannot be found to
better than a factor of four, for a 10 ms rotation period.  For two 10
$M_\odot$ black holes, the error in the net ``Kerr parameter per
mass'' of the
system projected orthogonal to the orbital plane,
$\Delta (a_1/m_1 + a_2/m_2)_z$, can be estimated to be about $\pm
0.1$.  However, Monte-Carlo studies indicate that
strong correlations between post-Newtonian terms
(dependent on $\mu$) and spin-orbit terms are likely to weaken these
estimates substantially \cite{jugger}.

Spin-spin terms, on the other hand, have negligible effect on
the accumulated phase for most systems of interest.  Only for two
extreme Kerr black holes are these terms discernable; for two $10
M_\odot$ black holes with both spins
orthogonal to the orbital plane, we estimate $\Delta (S_1\,S_2)
/\vert S_1S_2 \vert \simeq 0.4$.

\acknowledgments

We are grateful to Kip Thorne for sharing his early calculations of
spin effects with us and for encouraging us to pursue this problem
vigorously.  We also acknowledge useful discussions with Sam Finn.
This research is supported in part by the
National Science Foundation Grant No. 89-22140.

\end{document}